\documentclass{kluwer}
\usepackage{graphicx}

\begin{document}

\begin{article}
\begin{opening}
\title{Magnetic Reconnection and Turbulent Mixing: From ISM to Clusters of Galaxies} 
           
\author{\surname{Lazarian, A.}}
\institute{Dept. of Astronomy, UW-Madison, \email{lazarian@astro.wisc.edu}}
\author{\surname{Cho, J.}}
\institute{Dept. of Astronomy, UW-Madison, \email{cho@astro.wisc.edu}}

\runningtitle{Magnetic Reconnection and Mixing}
\runningauthor{Lazarian \& Cho}

\begin{abstract} 
Magnetic reconnection, or the ability of the magnetic field lines that are 
frozen in plasma to change their topology, 
is a fundamental problem of magnetohydrodynamics (MHD). We
briefly examine the problem starting with the well-known Sweet-Parker scheme, 
discuss effects
of tearing modes, anomalous resistivity and the concept of hyperresistivity. 
We show that the field stochasticity by itself provides a way to
enable fast reconnection even if, at the scale of individual turbulent wiggles,
the reconnection happens at the slow Sweet-Parker rate. 
We show that fast reconnection
allows efficient mixing of magnetic field in the direction perpendicular to
the local direction of magnetic field. 
While the idea of stochastic reconnection still requires
numerical confirmation,
our numerical simulations 
testify that mixing motions perpendicular to the local magnetic field are up
to high degree hydrodynamical. 
This suggests that the turbulent heat
transport should be similar to that in non-magnetized turbulent fluid, namely,
should have a diffusion coefficient $\sim V_L L$, where $V_L$ is the amplitude
of the turbulent velocity and $L$ is the scale of the turbulent motions. We
present numerical simulations which support this conclusion. The application
of this idea to thermal conductivity in clusters of galaxies shows that this
mechanism may dominate the diffusion of heat and may be efficient enough
to prevent cooling flow formation. 
\end{abstract}


\end{opening}

\section{Introduction}

Interstellar plasma has high conductivity and therefore magnetic field and fluid move together.
As the result magnetic field influences fundamental properties of interstellar plasma and
must be accounted in star formation processes, support and evolution of molecular clouds,
transfer of mass and energy in the interstellar space etc. Magnetic fields control 
cosmic rays propagation, induce Parker (see Parker 1979) and magneto-rotational 
(see Balbus \& Hawley 1998) instabilities. Transport processes in the interstellar
medium and beyond it, e.g. clusters of galaxies, depend on the ability of magnetic
field lines to change their topology or to reconnect. It is difficult to overestimate
the role of magnetic reconnection as it is
also a necessary component of magnetic dynamo 
(see Parker 1979, and also Vishniac, Lazarian \& Cho 2003 for a recent review),
and it is likely to affect magnetic flux removal from hotbeds of star formation etc.
More generally, 
it is impossible to claim that we understand MHD unless
we can predict whether crossing magnetic flux tubes will reconnect or bounce
from one another. 

Mixing motions perpendicular to magnetic field lines are an essential part of the
Goldreich-Sridhar model (1995, henceforth GS95) of MHD turbulence. Should those motions proceed
in the hydrodynamic-type fashion or magnetic field would accumulate unresolvable
knots that would affect the motion of the fluid? The answer to this question has
far reaching consequences for heat transport in the interstellar medium and beyond.
Indeed, magnetic fields are known to suppress heat conductivity 
(see Chandran \& Cowley 1998)
in the direction perpendicular to ${\bf B}$. Transport of heat along wondering
magnetic field lines (Narayan \& Medvedev 2001, Zakamska \& Narayan 2002) 
partially alleviates the problem. 
If reconnection proceeds efficiently, the mixing motions
should allow an efficient transport of heat in magnetized turbulent fluids. 
This would
have big implications for problems from cooling flows 
in the clusters of galaxies
(see Fabian 1994) to mixing layers in our and other galaxies 
(see Slavin, Shull, \& Begelman 1993).

In \S 2 and \S3, we discuss reconnection and, in \S4 and \S5,
we consider turbulent mixing of magnetized plasma.

\section{Sweet-Parker, Petschek  Models and their Modifications}

The literature on magnetic reconnection is rich and vast (see, for
example, Priest \& Forbes (2000) 
and references therein). We start by discussing
a robust scheme proposed by Sweet and Parker (Parker 1957; Sweet 1958). 
In this scheme oppositely directed magnetic fields are brought
into contact over a region of length $L_x$ (see Fig.~1). 
In general there will be a shared component, of the same order as
the reversed component.  However, this has only a minor effect
on our discussion.  The gradient
in the magnetic field is confined to the current sheet, a region
of vertical size $\Delta$, within which the magnetic
field evolves resistively. The velocity of reconnection, $V_r$,
is the speed with which magnetic field lines enter the current
sheet, and is roughly $\eta\approx V_r \Delta$.
Arbitrarily high values of $V_r$ can be achieved (transiently) by
decreasing $\Delta$. However, for sustained reconnection there is
an additional constraint imposed by mass conservation.
The plasma initially entrained on the magnetic field
lines must escape from the reconnection zone. In the Sweet-Parker
scheme this means a bulk outflow, parallel to the field lines, 
within the current sheet.  Since the mass enters along a zone of width
$L_x$, and is ejected within a zone of width $\Delta$, this implies
\begin{equation}
\rho V_{rec} L_x = \rho' V_A \Delta~~~,
\label{mass_con}
\end{equation}
where we have assumed that the outflow occurs at the Alfv\'en velocity $V_A$.
This is actually an upper limit set by energy conservation.
If we ignore the effects of compressibility $\rho=\rho'$ and the
resulting reconnection velocity allowed by Ohmic diffusivity
and the mass constraint is 
\begin{equation}
V_{rec, sweet-parker}\approx V_A {\cal R}_L^{-1/2}, 
\end{equation}
where
${\cal R}_L$ is the Lundquist number using the current sheet {\it length}.
Depending on the specific astrophysical context, this gives a reconnection
speed which lies somewhere between $10^{-3}$ (stars) and $10^{-10}$ (the
galaxy) times $V_A$.

Attempts to accelerate Sweet-Parker reconnection are numerous.
We start by considering schemes to broaden the current sheet.
Anomalous resistivity is known to broaden current sheets in laboratory
plasmas. But this is microphysical broadening which is not important
when the thickness of the current sheet that in the Sweet-Parker scheme
scales as $L_x^{1/2}$ is much larger than the Larmor radius of the
thermally moving proton. The latter scale is of the order of
$200$~km in the interstellar medium and it would be naive to expect
to be able to squeeze the interstellar gas out through such a narrow slot (see
eq~(\ref{mass_con}).

Tearing modes are a robust instability connected to the
appearance of narrow current sheets (Furth, Killeen, \& Rosenbluth 1963).
Their importance for reconnection by many authors (see Strauss 1988, Zweibel 1989).
The tearing modes broaden the reconnection layer and enhance the reconnection
speed.  
A treatment of the problem in  Lazarian \& Vishniac (1999; hereafter LV99)
of 3D reconnection involving tearing modes
provided an estimate 
\begin{equation}
V_{rec, tearing}=V_A\left({\eta\over V_A L_x}\right)^{3/10},
\label{tearing}
\end{equation}
which is substantially faster than the Sweet-Parker rate, but still
very slow in any astrophysical context. Note that unlike anomalous
effects, tearing modes do not require any special conditions and therefore
should constitute a generic scheme of reconnection.

Hyperresistivity was introduced in the literature (see Hameiri
\& Bhattacharjee 1987, Strauss 1988) to describe the 
ability of current sheet instabilities to drive 
plasma turbulence that would affect the reconnection rates. Tearing instability
is one of the major current sheet instabilities and therefore the concept of
hyperresistivity is related to the previously discuss issue. If the reconnecting
magnetic fields are nearly anti-parallel, the turbulent mixing of magnetic field
lines becomes possible and the reconnection can go much faster. However, this
assumption of very nearly parallel field lines substantially limits the application
of the concept.

The failure to find fast reconnection speeds due to 
broadening the current sheet 
has stimulated interest  to 
fast reconnection through radically different global 
geometries.  
Petschek (1964)
conjectured that reconnecting magnetic
fields would tend to form structures whose typical size in
all directions is determined by the resistivity (`X-point' reconnection).
This results in
a reconnection speed of order $V_A/\ln {\cal R}_L$.  However,
attempts to produce such structures in numerical simulations
of reconnection have been disappointing.  Typically
the X-point region collapses toward the
Sweet-Parker geometry as the Lundquist number becomes
large (Biskamp 1984, 1986, 1996; Wang, Ma, \& Bhattacharjee 1996;
Ma \& Bhattacharjee 1996).\footnote{Recent plasma reconnection experiments 
(Yamada et al.~2000)
do not support Petschek scheme either.}

One may invoke collisionless plasma effects
to stabilize the X-point reconnection (for collisionless plasma).
For instance, a number of authors 
(Shay et al.~1998; Shay \& Drake 1998; Shay et al.~1999)
have reported that in a two fluid treatment of magnetic reconnection,
a standing whistler mode can stabilize an X-point with a scale comparable
to the ion plasma skin depth, $c/\omega_{pi}\sim (V_A/c_s)r_L$.
The resulting reconnection speed is a large fraction of $V_A$, and
apparently 
independent of $L_x$, which would suggest that something like
Petschek reconnection emerges in the collisionless regime. 
However,
these studies have not yet demonstrated the possibility of fast reconnection
for generic field geometries, since they assume that there are no
bulk forces acting to produce a large scale current
sheet. Similarly, those studies do not account for fluid turbulence. 
Magnetic fields embedded in a turbulent fluid will give
fluctuating boundary conditions for the current sheets.
On the other hand,  boundary conditions need to be fine tuned for a
Petschek reconnection scheme (Priest \& Forbes 2000).

\section{Stochastic Reconnection}

The scheme of the stochastic reconnection is presented in Fig.~1.
We consider the case in which there exists a large scale,
well-ordered magnetic field, of the kind that is normally used as
a starting point for discussions of reconnection.  This field may,
or may not, be ordered on the largest conceivable scales.  However,
we will consider scales smaller than the typical radius of curvature
of the magnetic field lines, or alternatively, scales below the peak
in the power spectrum of the magnetic field, so that the direction
of the unperturbed magnetic field is a reasonably well defined concept.
In addition, we expect that the field has some small scale `wandering' of
the field lines.  On any given scale the typical angle by which field
lines differ from their neighbors is $\phi\ll1$, and this angle persists
for a distance along the field lines $\lambda_{\|}$ with
a correlation distance $\lambda_{\perp}$ across field lines (see Fig.~1).
Unlike the traditional Sweet-Parker scheme it allows many magnetic 
field lines to enter simultaneously into the reconnection region.
The reconnection therefore happens simultaneously at many surfaces
and over substantially smaller scales. Therefore the reconnection
velocity increases substantially. The limitation for the scheme
is the outflow of the matter, which in this case is controlled
by magnetic field line diffusivity. More discussion of the scheme
is provided below.

Note that the model of stochastic reconnection is 3 
dimensional\footnote{The Sweet-Parker scheme can
easily be extended into three dimensions, in the sense that one can take
a cross-section of the reconnection region such that the shared
component of the two magnetic fields is perpendicular to the
cross-section.  
In terms of the mathematics nothing changes, but the outflow velocity
becomes a fraction of the total $V_A$ and the shared component of the
magnetic field will have to be ejected together with the plasma. 
This result has motivated
researchers to do most of their calculations in 2D, which has obvious
advantages for both analytical and numerical investigations.}.  Our
 result cannot be obtained by considering two dimensional
turbulent reconnection (cf. Matthaeus \& Lamkin 1985).
In fact, it
does not arise from the turbulent transport of magnetic flux, 
as it incorrectly understood in Kim \& Diamond (2001),
but is a geometric effect arising from the appearance of stochastic
field line wandering in three dimensions.

The modification of the mass conservation constraint in the presence of
a stochastic magnetic field component 
is self-evident. Instead of being squeezed from a layer whose
width is determined by Ohmic diffusion, the plasma may diffuse
through a much broader layer, $L_y\sim \langle y^2\rangle^{1/2}$ (see Fig.~1),
determined by the diffusion of magnetic field lines.  This suggests
an upper limit on the reconnection speed of 
\begin{equation}
V_{rec,~mass~constr}\sim V_A (\langle y^2\rangle^{1/2}/L_x)
\label{mass}
\end{equation}

\begin{figure}
\begin{center}
\includegraphics[width=0.60\textwidth]{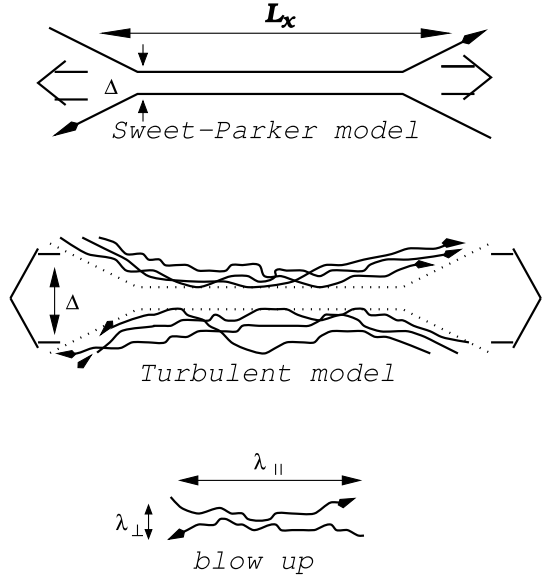}
\end{center}
 \caption[]{Upper plot: Sweet-Parker scheme of reconnection. Middle plot:
scheme of stochastic reconnection that accounts for field line
stochasticity.  Lower plot: a blow up of the contact region. 
Thick arrows depict outflows of plasma. From Lazarian \& Vishniac (2000)}
\end{figure}

To make further progress one should adopt a model of magnetic turbulence.
LV99 considered the whole range of possible models.
Using GS95 scalings and assuming that the localized reconnection goes
at the Sweet-Parker rate it is possible to get
\begin{equation}
V_{rec,~ SP-Alfven,~ local}\sim V_A (\eta/V_A L)^{1/4}
\end{equation}
where the turbulent velocity at the scale $L$ is assumed to be Alfvenic.
The isotropy of the fast modes has been proved numerically 
(Cho \& Lazarian 2002, 
also review by Cho \& Lazarian 2003). Our work suggests that
both in magnetic pressure dominated and gas pressure dominated environments
fast modes scaling is similar to that of acoustic turbulence. 
Assuming that the spectrum of such waves is truncated because of resistivity,
LV99
obtained
\begin{equation}
V_{rec,~SP-fast,~local}\sim V_A (\eta/V_A L)^{1/6},
\end{equation}
which is faster than the Alfvenic turbulence can provide.

If we assume that resistivity presents the bottleneck for the reconnection rate,
the global reconnection rate can be estimated as 
$V_{rec~local}L_x/\lambda_{\|}$.
If the plasma is fully ionized $\lambda$ gets very small and the limit
for the global reconnection rate provided by the resistivity
scales as $V_A Rm^{1/4}\gg V_A$ for Alfven modes and $V_A Rm^{1/2}\gg V_A$ for 
fast modes. 
Similarly, considering other possible bottlenecks, LV99 concluded that the mass
conservation (eq.~(\ref{mass})) presents the most stringent constraint 
on the reconnection
rate for the reconnection of the fully ionized gas. 
For Alfvenic turbulence this gives (see LV99)
\begin{equation}
V_{rec, global}\approx V_A min\left[(L_x/l)^{1/2}, (l/L_x)^{1/2}\right] (V_L/V_A)^2
\end{equation}
The case of the partially ionized
gas is considered in Lazarian, Vishniac, \& Cho (2003). 
There it is found that in
the assumption of the local Sweet-Parker rates, 
the reconnection is slower because
the turbulent cascade gets modified by ambipolar drag. 
However, the resulting
reconnection rates are sufficiently high to be important for removing flux
from star-forming clouds.

\section{Mixing and Turbulent Diffusion in Magnetized Media}

Let us start with a  note of warning. The scheme of stochastic reconnection
that we have discussed is very
different from the so-called ``turbulent diffusivity of magnetic field''.
The latter is a wrong concept that was used
at earlier stages of the mean dynamo theory (see recent review
by Vishniac, Lazarian \& Cho 2003) to justify some of its assumptions.
According to ``turbulent diffusivity'' idea, the magnetic fields of
opposite polarity can be mixed up by turbulence and then dissipate at
small scales. This picture is wrong as magnetic fields are dynamically
important and they get to dominate the small scale motions of the
magnetized fluid (see Parker 1992). 
In the stochastic reconnection scheme magnetic field wondering
happens within each layer of the opposite polarity, but no
small scale mixing of the magnetic field of opposite polarities is assumed.

Mixing of magnetic field lines happens in the direction perpendicular
to local direction of the magnetic field. GS95 picture of turbulence
can be understood in terms of the eddies in the planes
perpendicular to magnetic 
field lines (see discussion in Cho, Lazarian \& Vishniac 2002).
LV99 showed that the eddies will not be forming magnetic knots if
the reconnection is as fast as the stochastic reconnection scheme  
suggests. This means that the motions of the magnetized fluid will
be very similar to the hydrodynamic motions in the planes
perpendicular to the local direction of magnetic field.

Cho, Lazarian \& Vishniac (2002) scrutinized the
statistics of the motions of the magnetized turbulent
fluid in the direction perpendicular to the local direction
of magnetic field. Using numerical simulations they found that for sufficiently strong
magnetic field the turbulent motions perpendicular to magnetic field are
identical to the hydrodynamic motions. 
{\it Although this
fact cannot be used as a proof of the stochastic reconnection
but it can be used as the evidence that, in the
absence of any anomalous effects, the change of
magnetic topology does not constrain fluid motions.}

While the ``turbulent diffusivity of magnetic field'' is
a faulty concept, turbulence may provide efficient diffusivity
for heat and matter. Indeed, the turbulent motions perpendicular
to magnetic field should transport heat and other passive scalar
field in the same way as the hydrodynamic turbulence does. 
Below we will be mostly talking about heat transfer, as
this is the problem of largest astrophysical significance.
The corresponding diffusion coefficient is expected to be of
the order of $L V_L$.  

Do we expect the turbulent heat transport to be
strongly anisotropic?
As the large scale field wanders, the
direction of the local magnetic field that determines
the mixing motions of hydrodynamic-like eddies changes
as well. Therefore no strong anisotropy in heat conduction
is expected.

To test our theoretical considerations related
to the diffusion of heat we performed numerical
simulations using our hybrid ENO MHD code described, for instance, in
Cho \& Lazarian (2002). 
We use a passive scalar $\psi ({\bf x})$ to trace thermal particles.
We inject a passive scalar with a Gaussian profile:
\begin{equation}
   \psi({\bf x},t=t_0) \propto \exp^{-({\bf x}-{\bf x}_0)^2/\sigma_0^2}, 
   \label{eq_dist}
\end{equation}
where $\sigma_0$= $1/16$ of a side of the numerical box and ${\bf x}_0$ 
lies at the center of the computational box.
The value of $\sigma_0$ ensures that the scalar is injected
in the inertial range of turbulence.
The scalar field follows the continuity equation:
${\partial \psi    }/{\partial t} + \nabla \cdot (\psi {\bf v}) =0$.

\begin{figure*}
  \includegraphics[width=0.33\textwidth]{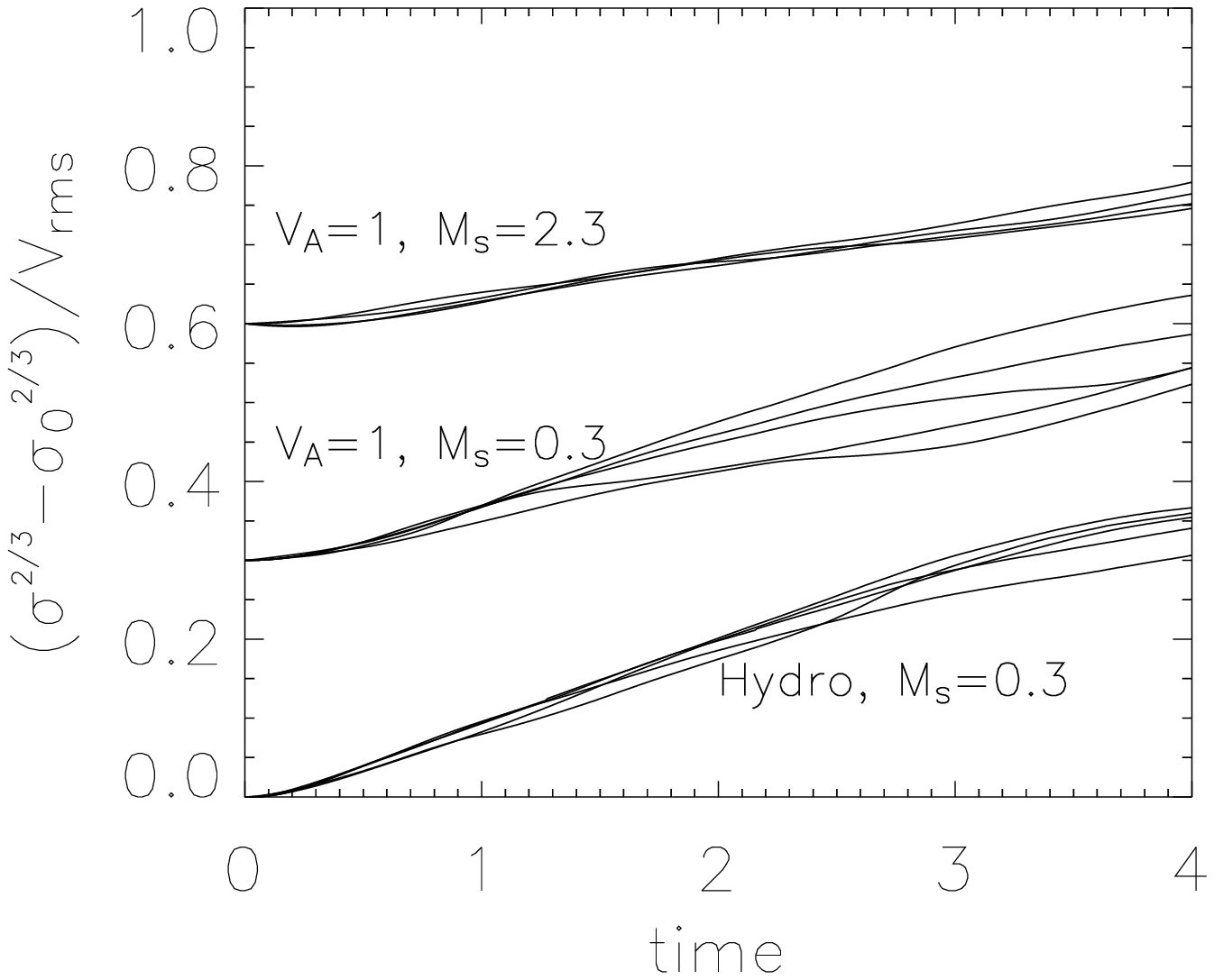}
\hfill
  \includegraphics[width=0.32\textwidth]{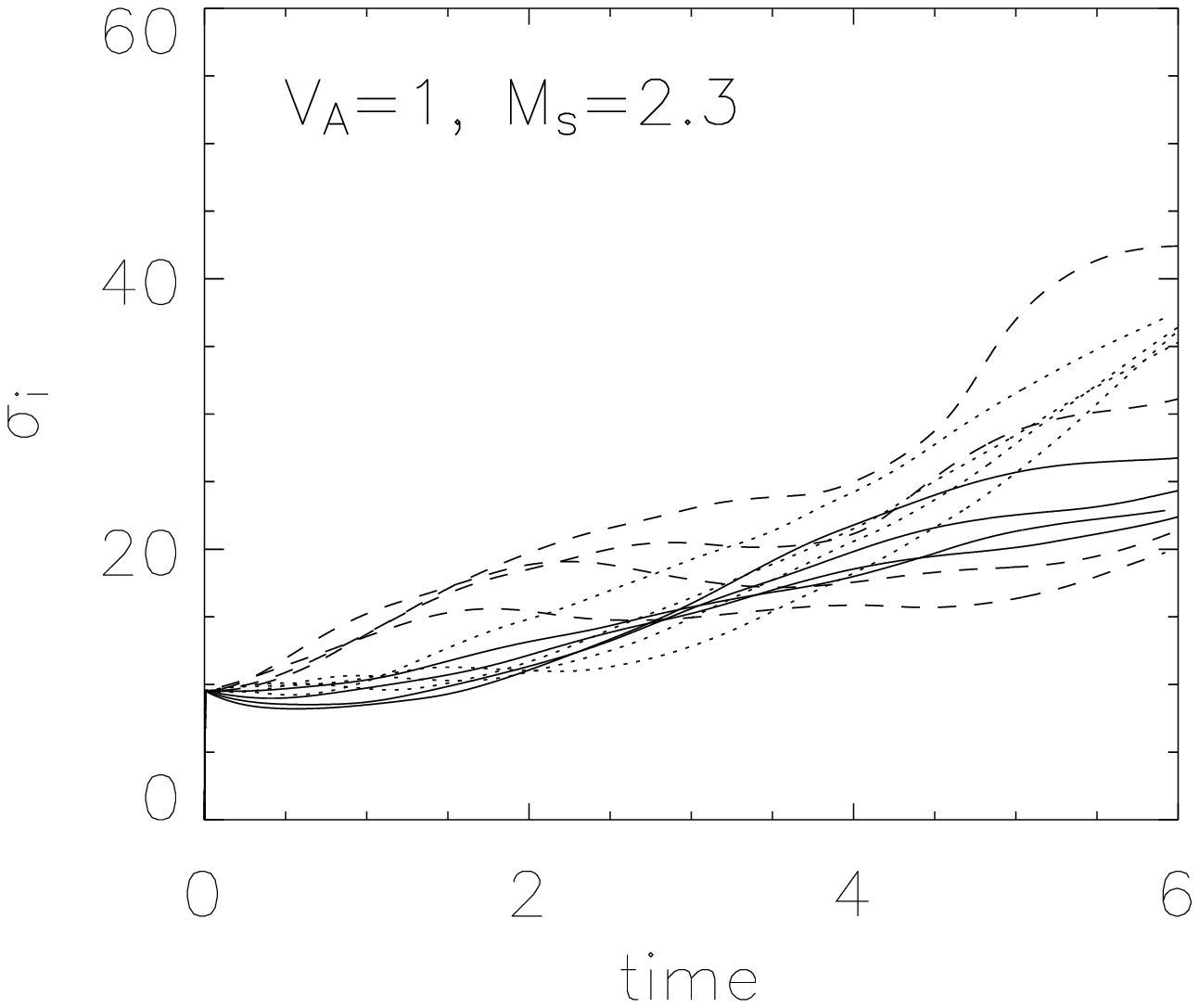}
\hfill
  \includegraphics[width=0.28\textwidth]{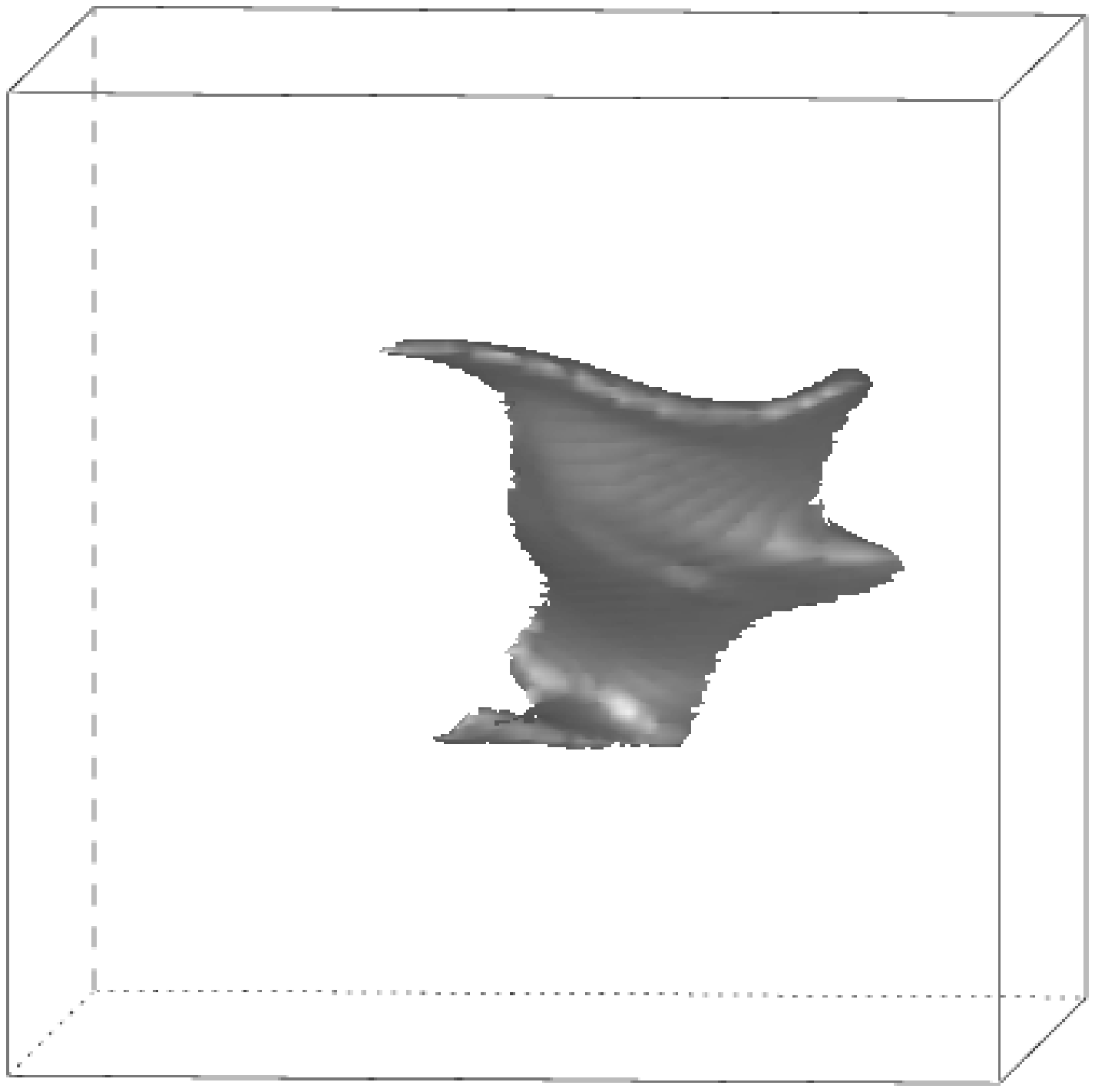}
  \caption{
   {\it Left:} $(\sigma^{2/3}-\sigma_0^{2/3})/V_{L}$ vs. time.
   $\sigma$ is the standard deviation of the passive scalar field.
   Y-axis is in the unit of box-size and
   Y-values are shifted by 0.3 units for convenience.
   Note that the slope does not strongly depend on the mean field $B_0$ or 
   sonic Mach number $M_s$.
   {\it Middle:} $\sigma_i$ (i=x, y, and z) vs. time.
   Solid lines=parallel to ${\bf B}_0$; dashed and dotted
   lines=perpendicular to
   ${\bf B}_0$.
   {\it Right:} Snapshot of the passive scalar field for 
   $V_A (\equiv B_0/\sqrt{4 \pi \rho}) \sim V_L$ and
   $M_s=0.3$. ${\bf B}_0$ is parallel to the dashed line (i.e. vertical).
   Turbulent motions provide efficient mixing of the passive scalar.
   {\it Left} and {\it Middle} panels suggest that
   magnetic field does not suppress this turbulent mixing process.
   Obviously, when $V_A$ is lower than $V_L$, the efficiency of
   turbulent mixing is no less than that in the $V_A\sim V_L$ case.
   Results are obtained from using $192^3$ and $216^3$ grid points.
}
\end{figure*}

In Figure~2,
we compare time evolution of $\sigma$ in hydrodynamic case and in
MHD cases. The results are very close and this enables us to 
suggest that the coefficient of thermal diffusion is 
\begin{equation}
 \kappa_{dynamic} = C_{dyn} L V_L,
\end{equation}
where $C_{dyn}$ is a constant of order unity, $V_L$ is the amplitude
of the r.m.s. turbulent velocity, and $L$ is the scale of the turbulent motions.
This is the effective diffusion by turbulent motions suitable for scales
larger than injection scale $L$.
The value of $C_{dyn}$ remains almost constant for 
$V_A$'s of up to the equipartition value
$V_A  \sim V_L$.
The exact value of $C_{dyn}$ is uncertain.
In hydrodynamic cases, $C_{dyn}$ is of order of $\sim 0.3$ (see Lesieur 1990
chapter VIII and references therein). 

\section{Astrophysical Implications}

In the sections above we discussed two issues: magnetic stochastic
reconnection and turbulent transport of heat through mixing motions.
We believe that the two processes are intrinsically connected and fast reconnection 
implies efficient turbulent transport. Nevertheless, while testing of
stochastic reconnection is still a challenging problem, efficient mixing
and heat transport has been already tested numerically. 

The astrophysical implications of stochastic reconnection were discussed in LV99
where it was shown that it naturally accounts for Solar flares, 
for operation of galactic dynamo, etc. In Lazarian
et al. (2003), its implications for gamma ray busts were analyzed, however.
This does not limit the list of the possible applications. 
For instance, it is easy to see that stochastic reconnection can provide efficient removal
of magnetic flux from starforming clouds. 
A proper discussion of these implications
requires a more extended review. 
Here we briefly consider implications of fast turbulent transport.

\vspace{0.3cm}
{\bf Clusters of Galaxies}\\
It is widely
accepted that ubiquitous X-ray emission due to hot gas in clusters of
galaxies should cool significant amounts of intracluster medium and
this must result in cooling flows (Fabian 1994). However, observations
do not support the evidence for the cool gas (see Fabian et al. 2001)
which is suggestive of the existence of heating that replenishes the 
energy lost via X-ray emission. Heat transfer from the outer hot regions
can do the job, provided that the heat transfer is sufficiently efficient.

Gas in the clusters of galaxies is magnetized and the conventional wisdom
maintains that the magnetic fields strongly suppress thermal conduction
perpendicular to their direction. Realistic magnetic fields are turbulent
and the issue of the thermal conduction in such a situation has been
long debated. A recent paper by Narayan \& Medvedev (2001) obtained the
estimates for the thermal conductivity of the turbulent magnetic fields.

However, Narayan \& Medvedev (2001) treated the turbulent magnetic
fields as static. In hydrodynamical turbulence
it is possible to neglect plasma turbulent motions only when the diffusion of
electrons which is the product of the electron thermal velocity $v_{elect}$
and the electron mean free path in plasma $l_{mfp}$, i.e. 
$v_{elect}l_{mfp}$, is greater than the turbulent velocity $V_L$
times the turbulent injection scale $L$, i.e. $LV_L$.
If such scaling estimates are applicable to heat transport in magnetized
plasma, the turbulent heat transport should be accounted for heat transfer
within clusters of galaxies. Indeed, data for  $l_{mfp}$
$v_{elect}l_{mfp}$ given in Zakamska \& Narayan (Narayan \& Medvedev 2001)
provide the diffusion 
coefficient
$\kappa_{Sp}\equiv v_{elect}l_{mfp} \sim 6.22 \times 10^{30}$ cm$^2$ sec$^{-1}$
(for Hydra A). 
If turbulence in the cluster of galaxies is of the order of
the velocity dispersion of galaxies, while the injection scale $L$ is
of the order of $20$~kpc, the turbulent diffusion coefficient is 
$\kappa_{dynamic}\sim LV_L
\sim 6 \times 10^{30}$ cm$^2$ sec$^{-1}$, where we take
$V_L \sim 1000$ km/sec.

Although the estimates for the electron thermal diffusivity and our
estimate of the turbulent diffusivity are of the same order
the applicability of Narayan \& Medvedev's model is a bit restricted - 
their model requires strong (i.e. $B_0 \sim \delta B$) mean magnetic field.
While there are strong mean magnetic fields in the Galaxy, this is unlikely
for the intracluster medium.
When the mean field is weak, turbulence at
the scales smaller than the characteristic
magnetic field scale ($\equiv l_{B}$) 
may follow the GS95 model.
However, this requires further studies.
Our turbulent mixing model  gives the same $\kappa_{dynamic}$
regardless of magnetic field geometry.

\vspace{0.3cm}
{\bf Local Bubble and SNRs}\\
Now let us compare the results for thermal diffusivity for
other important astrophysical situations.
The Local Bubble (LB) is a hot ($T\sim 10^6K; kT\sim$ 100 eV), 
tenuous ($n\sim 0.008/cm^3$) cavity
immersed in the interstellar medium (Berghofer et al 1998; Smith \& Cox 2001).
Turbulence parameters are uncertain.
We take typical interstellar medium values: $L\sim$ 10 pc and
$V_L\sim$ 5 km/sec.
For these parameters, the ratios of $\kappa_{dynamic}$ to $\kappa_{Sp}$ are
\begin{eqnarray}
    \mu_{in} & = &      \kappa_{dynamic} /\kappa_{Sp} \sim  0.05,
           \mbox{~~~(inside LB)} \\
    \mu_{mix} & = &       \kappa_{dynamic} /\kappa_{Sp} \sim  100,
           \mbox{~~~(in mixing layer),}
\end{eqnarray}
where we take $\bar{T}\sim \sqrt{T_cT_h}\sim 10^5 K$, 
$\bar{n}\sim \sqrt{n_cn_h}\sim 0.1/cm^3$ (Begelman \& Fabian 1990),
$T_c\sim 10^4 K$, $n_c \sim 1/cm^3$, 
$T_h\sim 10^6 K$, and $n_h \sim 0.008/cm^3$.
Subscripts `h' and `c' stand for `hot' and `cold', respectively.
This implies that 
thermal conduction along static magnetic field lines is
suppressed when electrons gyrate through the `cold' and `dense' mixing layer. 
Note that turbulent diffusion is still effective in the mixing layer.
We expect similar results for supernova remnants 
since parameters are similar.

\section{Summary}

In the article above we briefly reviewed a number of existing ideas
about reconnection. Analysis of the solar and other reconnection-related data
provides a number of requirements to the successful candidate. It should
be robust to act in the situation when the boundary conditions constantly
change due to turbulence. It should typically provide reconnection rates
of the order of one tenth $V_A$, but sometimes it should be much slower
to allow the accumulation of the magnetic flux. Our analysis shows
that the stochastic reconnection satisfies naturally to all these criteria.

Stochastic reconnection is different from the ``turbulent magnetic 
diffusivity'' idea. It does not require fine mixing of the magnetic
fields of opposite polarity. However, we show that the turbulent transport of heat
happens in the presence of magnetic field very similarly to how it happens in
pure hydrodynamics. In particular, we show that
turbulent transport of heat in the intracluster matter is at least as fast as
the earlier estimates based on the electron thermal conductivity suggested.

\acknowledgements
A.L. acknowledges  the support of NSF Grant AST-0125544.
The study of turbulent diffusion was done in collaboration with
P. Moin and A. Honein at the Center for Turbulence Research 
in Stanford University as a part of Visiting Scholar Program in Summer 2002.

\end{article}
\end{document}